# Mobility and Saturation Velocity in Graphene on SiO$_2$

Vincent E. Dorgan,[1] Myung-Ho Bae,[1] and Eric Pop[1, 2, *]

[1]*Dept. of Electrical and Computer Engineering, Micro and Nanotechnology Laboratory, University of Illinois, Urbana-Champaign, IL 61801, USA*
[2]*Beckman Institute, University of Illinois, Urbana-Champaign, IL 61801, USA*

We examine mobility and saturation velocity in graphene on SiO$_2$ above room temperature (300-500 K) and at high fields (~1 V/µm). Data are analyzed with practical models including gated carriers, thermal generation, "puddle" charge, and Joule heating. Both mobility and saturation velocity decrease with rising temperature above 300 K, and with rising carrier density above $2\times10^{12}$ cm$^{-2}$. Saturation velocity is $>3\times10^7$ cm/s at low carrier density, and remains greater than in Si up to $1.2\times10^{13}$ cm$^{-2}$. Transport appears primarily limited by the SiO$_2$ substrate, but results suggest intrinsic graphene saturation velocity could be more than twice that observed here.

[*]Contact: epop@illinois.edu

The excellent electrical and thermal properties of graphene hold great promise for applications in future integrated-circuit technology.[1] For instance, the electron and hole energy bands are symmetric,[1,2] leading to equal and high electron and hole mobilities, unlike in typical semiconductors like Si, Ge or GaAs where hole mobility is lower. However, despite many measurements at low fields and low temperatures,[3] surprisingly little data or models exist for transport in graphene at temperatures and high electric fields typical of modern transistors.

In this study we measure mobility in the $T$ = 300-500 K range and velocity saturation at fields $F \sim 1$ V/μm in monolayer graphene on SiO$_2$, both as a function of carrier density. We also introduce simple models including the proper electrostatics, and self-heating[4] at high fields. We find that mobility and saturation velocity decrease with rising temperature above 300 K, and with rising carrier density above 2×10$^{12}$ cm$^{-2}$, and appear limited by the SiO$_2$ substrate. The relatively straightforward approach presented can be used for device simulations or extended to graphene on other substrates.

We fabricated four-probe graphene structures on SiO$_2$ with a highly doped Si substrate as the back-gate (Fig. 1a and supplementary material[5]). To obtain mobility and drift velocity from conductivity measurements, we model the carrier density including gate-induced ($n_{cv}$), thermally generated ($n_{th}$) carriers, electrostatic spatial inhomogeneity ($n^*$) and self-heating at high fields. Previous mobility estimates using only $n_{cv}$ could lead to unphysically high mobility (μ → ∞) near the Dirac voltage ($V_G = V_0$) at the minimum conductivity point.

First, we note the gate voltage imposes a charge balance relationship as

$$n_{cv} = p - n = -C_{ox}V_{G0} / q \qquad (1)$$

where $C_{ox} = \epsilon_{ox}/t_{ox}$ is the capacitance per unit area (quantum capacitance may be neglected here[6,7]), $\epsilon_{ox}$ is the dielectric constant of SiO$_2$, $q$ is the elementary charge, and $V_{G0} = V_G - V_0$ is the gate voltage referenced to the minimum conductivity point. Next, we define an average Fermi level $E_F$ such that $\eta = E_F/k_BT$, leading to the mass-action law:[6]

$$pn = n_{th}^2 \frac{\Im_1(\eta)\Im_1(-\eta)}{\Im_1^2(0)} \qquad (2)$$

where $n_{th} = (\pi/6)(k_BT/\hbar v_F)^2$ is the thermal carrier density, $v_F \approx 10^8$ cm/s is the Fermi velocity, and $\Im_j(\eta)$ is the Fermi-Dirac integral, $\Im_1(0) = \pi^2/12$.

Next, we account for the spatial charge ("puddle") inhomogeneity of graphene due to substrate impurities.[8,9] The surface potential can be approximated[7] as a periodic step function whose amplitude $\pm\Delta$ is related to the width of the minimum conductivity plateau,[5,10] as given by the residual carrier puddle density ($n^*$) due to charged impurities in the SiO$_2$ ($n_{imp}$). We find $n^* \approx 0.297 n_{imp} \approx 2.63 \times 10^{11}$ cm$^{-2}$ here,[5] i.e. a surface potential variation $\Delta \approx 59$ meV. This is similar to a previous study (~54 meV),[7] and to scanning tunneling microscopy results (~77 meV).[9] The surface potential inhomogeneity is equivalent to a Dirac voltage variation $\Delta V_0 = q n^*/C_{ox} \approx 3.66$ V.

The total carrier density can be determined numerically by averaging Eqs. (1) and (2) for the regions of $\pm\Delta$, but does not yield an analytic expression. In order to simplify this, we note that at low charge density ($\eta \rightarrow 0$) the factor $\mathfrak{I}_1(\eta)\mathfrak{I}_1(-\eta)/\mathfrak{I}_1^2(0)$ in Eq. (2) approaches unity. At large $|V_{G0}|$ the gate-induced charge dominates, i.e. $n_{cv} \gg n_{th}$ when $\eta \gg 1$. Finally, we add a correction for the spatial charge inhomogeneity which gives a minimum carrier density $n_0 = [(n^*/2)^2 + n_{th}^2]^{1/2}$ resulting from averaging the regions of $\pm\Delta$. Solving Eqs. (1) and (2) above with these approximations results in an *explicit* expression for the concentration of electrons and holes:

$$n, p \approx \frac{1}{2}\left[\pm n_{cv} + \sqrt{n_{cv}^2 + 4n_0^2}\right] \qquad (3)$$

where the lower (upper) sign corresponds to electrons (holes). Equation (3) can be readily used in device simulations and is similar to a previous empirical formula,[11] but derived here on rigorous grounds. We note Eq. (3) reduces to the familiar $n = C_{ox}V_{G0}/q$ at high gate voltage, and to $n = n_0$ (puddle regime) at $V_G \sim V_0$. Figure 1b displays the role of thermal and "puddle" corrections to the carrier density at 300 K and 500 K. These are particularly important near $V_{G0} = 0$ V, when the total charge density relevant in transport ($n+p$) approaches a constant despite the charge neutrality condition imposed by the gate ($n-p = 0$). At higher temperatures ($k_B T \gg \Delta$) the spatial potential variation becomes less important due to thermal smearing and higher $n_{th}$.

Using the above, mobility is obtained as $\mu = \sigma/q(p+n)$ at low fields (~2 mV/μm), where the conductivity $\sigma = (L/W)I_{14}/V_{23}$. Mobility is shown in Fig. 2a at various temperatures and $V_{G0} > 0$ ($n > p$), i.e. electron majority carriers.[12] (hole mobility and additional discussion is available in the supplement[5]). The mobility here peaks at ~4500 cm$^2$/V·s and decreases at carrier densities greater than ~2×10$^{12}$ cm$^{-2}$, at 300 K. Mobility decreases with rising $T > 300$ K for all carrier den-sities (Fig. 2b), as was also noted by Ref. 7 albeit in a lower temperature range. The dependence of mobility on carrier density and temperature suggests the dominant scattering mechanism changes from Coulomb to phonon scattering at higher densities and temperatures.[7] Inspired by empirical approximations for Si device mobility,[13] the data can be fit as (dashed lines in Fig. 2):[14]

$$\mu(n,T) = \frac{\mu_0}{1 + \left(n / n_{ref}\right)^\alpha} \times \frac{1}{1 + \left(T / T_{ref} - 1\right)^\beta} \tag{4}$$

where $\mu_0$ = 4650 cm$^2$/V·s, $n_{ref}$ = 1.1×10$^{13}$ cm$^{-2}$, $T_{ref}$ = 300 K, $\alpha$ = 2.2 and $\beta$ = 3.

We then turn our attention to high-field drift velocity measurements, which pose challenges due to Joule heating, non-uniform potential and carrier density along the channel. To account for self-heating we estimate the average device temperature via its thermal resistance $\mathcal{R}_{th}$ (Fig. 1a):[2]

$$\Delta T = T - T_0 \approx P\left(\mathcal{R}_B + \mathcal{R}_{ox} + \mathcal{R}_{Si}\right) \tag{5}$$

where $P = I_{14}V_{23}$, $\mathcal{R}_B = 1/(hA)$, $\mathcal{R}_{ox} = t_{ox}/(\kappa_{ox}A)$, and $\mathcal{R}_{Si} \approx 1/(2\kappa_{Si}A^{1/2})$ with $A = LW$ the area of the channel, $h \approx 10^8$ Wm$^{-2}$K$^{-1}$ the thermal conductance of the graphene-SiO$_2$ boundary,[15] $\kappa_{ox}$ and $\kappa_{Si}$ the thermal conductivities of SiO$_2$ and of the doped Si wafer.[5] At 300 K for our geometry $\mathcal{R}_{th}$ ≈ 10$^4$ K/W, or ~2.8×10$^{-7}$ m$^2$K/W per unit of device area. Of this, the thermal resistance of the 300 nm SiO$_2$ ($\mathcal{R}_{ox}$) accounts for ~84%, the spreading thermal resistance into the Si wafer ($\mathcal{R}_{Si}$) ~12% and the thermal resistance of the graphene-SiO$_2$ boundary ($\mathcal{R}_B$) ~4%. We note that the role of the latter two will be more pronounced for smaller devices on thinner oxides. The thermal model in Eq. (5) can be used when the sample dimensions are much greater than the SiO$_2$ thickness ($W, L \gg t_{ox}$) but much less than the Si wafer thickness.[2]

To minimize charge non-uniformity and temperature gradients along the channel at high field[4] we bias the device at high $|V_G|$ and avoid ambipolar transport ($V_{GS}–V_0$ and $V_{GD}–V_0$ have same sign).[11] We confirm this with finite-element simulations.[4,5] The drift velocity is $v = I_{14} /$ $(Wqn_{23})$ where $n_{23}$ is the average carrier density between terminals 2 and 3, and the background temperature is held at $T_0$ = 80 K and 300 K. Due to self-heating ($T = T_0 + \Delta T$), these enable measurements of saturation velocity ($v_{sat}$) near room temperature and above, respectively.

Figures 3a and 3b show the velocity-field relationship at the two background temperatures, indicating saturation tendency at fields $F > 1$ V/μm. We fit the drift velocity by

$$v(F) = \frac{\mu F}{\left[ 1 + \left( \mu F / v_{sat} \right)^\gamma \right]^{1/\gamma}} \tag{6}$$

where $\mu$ is the low-field mobility from Eq. (4) and $\gamma \approx 2$ provides a good fit for the carrier densities and temperatures in this work. To limit the role of self-heating, data was only fit up to $\Delta T \sim$ 200 K (solid symbols).[16] To ensure no sample degradation due to high field stress we checked that low-field $I$–$V_G$ characteristics were reproducible after each high bias measurement.[16]

Figure 3c shows extracted drift velocity vs. electron density (symbols) at $F = 2$ V/μm, for the two background temperatures. We compare these experimental results with an analytic model (dashed lines) which approximates the high-field distribution with the two half-disks shown in the Fig. 3c inset, suggested by previous simulations.[17] This model assumes $v_{sat}$ is limited by inelastic emission of optical phonons (OP) and leads to:[18]

$$v_{sat}(n, T) = \frac{2}{\pi} \frac{\omega_{OP}}{\sqrt{\pi n}} \sqrt{1 - \frac{\omega_{OP}^2}{4\pi n v_F^2}} \frac{1}{N_{OP} + 1} \tag{7}$$

where $\hbar\omega_{OP}$ is the OP energy and $N_{OP} = 1/[\exp(\hbar\omega_{OP}/k_B T) - 1]$ is the phonon occupation. At low temperature and low carrier density the result is a constant, $v_{max} = (2/\pi)v_F \approx 6.3 \times 10^7$ cm/s (six times higher than $v_{sat}$ in Si); at high carrier density it scales as $v_{sat} = (2/\pi)\omega_{OP}/(\pi n)^{1/2}$, dependent both on the OP energy and the carrier density $n$.

We consider two dominant phonon mechanisms in Fig. 3c, $\hbar\omega_{OP} = 55$ meV (lower dashed, SiO$_2$ substrate OP[19]) and $\hbar\omega_{OP} = 160$ meV (upper dashed, graphene OP[20]). The model limited by SiO$_2$ phonons slightly underestimates $v_{sat}$, while the model with graphene OPs significantly overestimates the measured $v_{sat}$. This suggests that both phonons play a role in limiting $v_{sat}$, but that substrate phonons are dominant for graphene on SiO$_2$. (For device simulations the fit can be optimized using an intermediate value $\hbar\omega_{OP} \approx 82$ meV.) Nevertheless, $v_{sat}$ is greater than in Si for charge densities $n < 1.2 \times 10^{13}$ cm$^{-2}$ and more than twice that of Si at $n < 4 \times 10^{12}$ cm$^{-2}$. With only the graphene OP ($\hbar\omega_{OP} = 160$ meV) the model suggests an upper limit for the "maximum" $v_{sat}$ that could be expected. This intrinsic $v_{sat}$ could be more than twice as high as that measured here on SiO$_2$ and from two to six times higher than in Si for the carrier densities considered here.

Finally, we note the data in Fig. 3c suggest a temperature dependence of $v_{sat}$, included here through the last term in Eq. (7). This term is qualitatively similar to that in Si,[21] and due to the OP scattering (emission) rate being proportional to $(N_{OP} + 1)$.[22] The model yields a ~20% decrease in $v_{sat}$ between ~280 K and ~500 K if the $SiO_2$ phonon is dominant, and a ~2% decrease if the graphene OP is dominant. The data in Fig. 3c show much closer agreement with the former, once again indicating the effect of the $SiO_2$ in limiting graphene transport.

In summary, we examined mobility and saturation velocity in graphene on $SiO_2$, including the roles of carrier density and temperature. The results focus on the $T > 300$ K and high field $F > 1$ V/µm regime, where few studies presently exist. Both data and models point to the effect of the $SiO_2$ substrate in limiting graphene transport. The models introduced are simple yet practical, and can be used in future simulations of graphene devices operating near room temperature and up to high fields.

We acknowledge financial support from the Nanoelectronics Research Initiative (NRI), ONR grant N00014-10-1-0061, NSF grant CCF 08-29907, and valuable discussions with S. Datta, D. Estrada, M. Fuhrer and E. Tutuc.

**FIGURES**

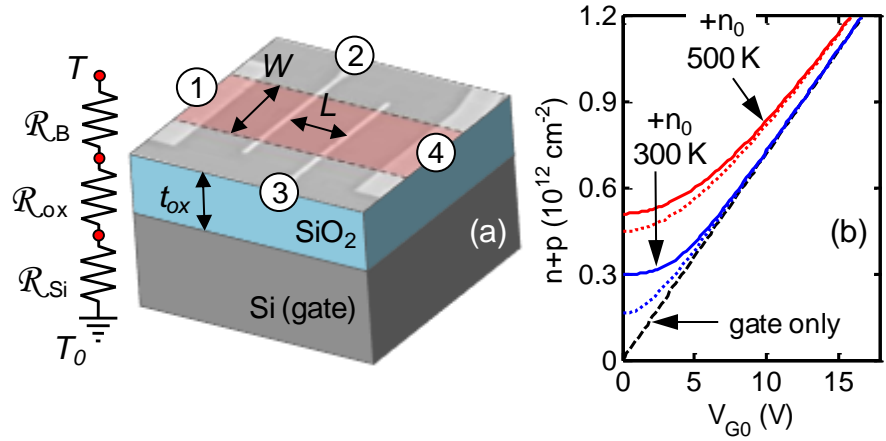

**FIG. 1 (a)** Schematic of graphene sample ($W$ = 7 μm, $L$ = 4 μm, $t_{ox}$ = 300 nm) connected to four-probe electrodes; graphene colorized for clarity. Thermal resistance model is used to calculate average temperature rise at high bias. **(b)** Calculated carrier density vs. gate voltage at 300 K (blue) and 500 K (red) in electron-doped regime ($n > p$). Solid lines include contribution from electrostatic inhomogeneity $n^*$ and thermal carriers $n_{th}$ (both relevant at 300 K), dotted lines include only $n_{th}$ (dominant at 500 K). Black dashed line shows only contribution from gating, $n_{cv}$.

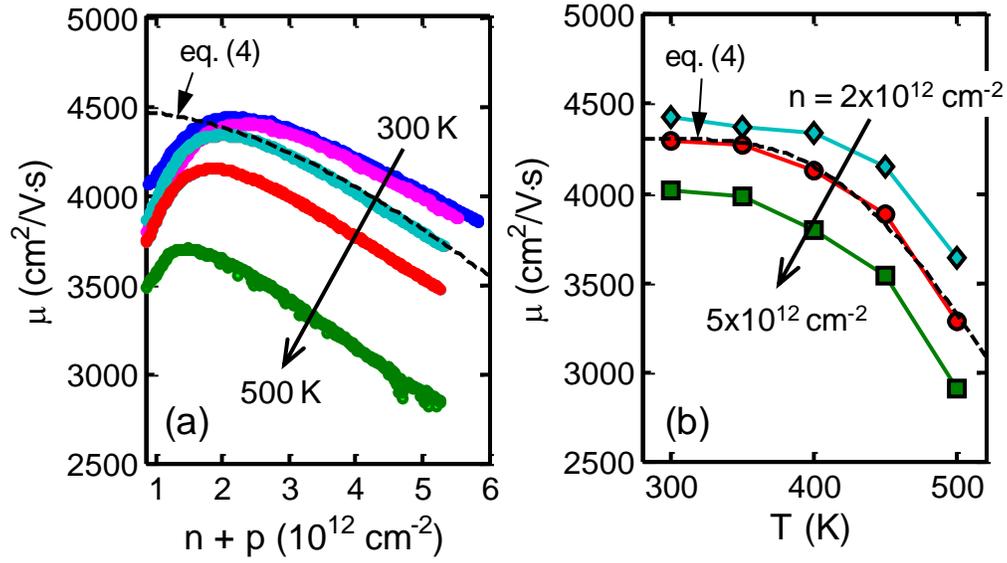

**FIG. 2 (a)** Mobility vs. carrier density in the electron-doped regime ($V_{G0} > 0$, $n > p$), obtained from conductivity measurements at $T = 300$–$500$ K, in 50 K intervals. The qualitative dependence on charge density is similar to that found in carbon nanotubes.[22] Dashed line shows fit of Eq. (4) with $T = 400$ K.[14] **(b)** Mobility vs. temperature at $n = 2 \times 10^{12}$ (top), $3.5 \times 10^{12}$ (middle), and $5 \times 10^{12}$ cm$^{-2}$ (bottom). Dashed line shows fit of Eq. (4) with $n = 3.5 \times 10^{12}$ cm$^{-2}$.

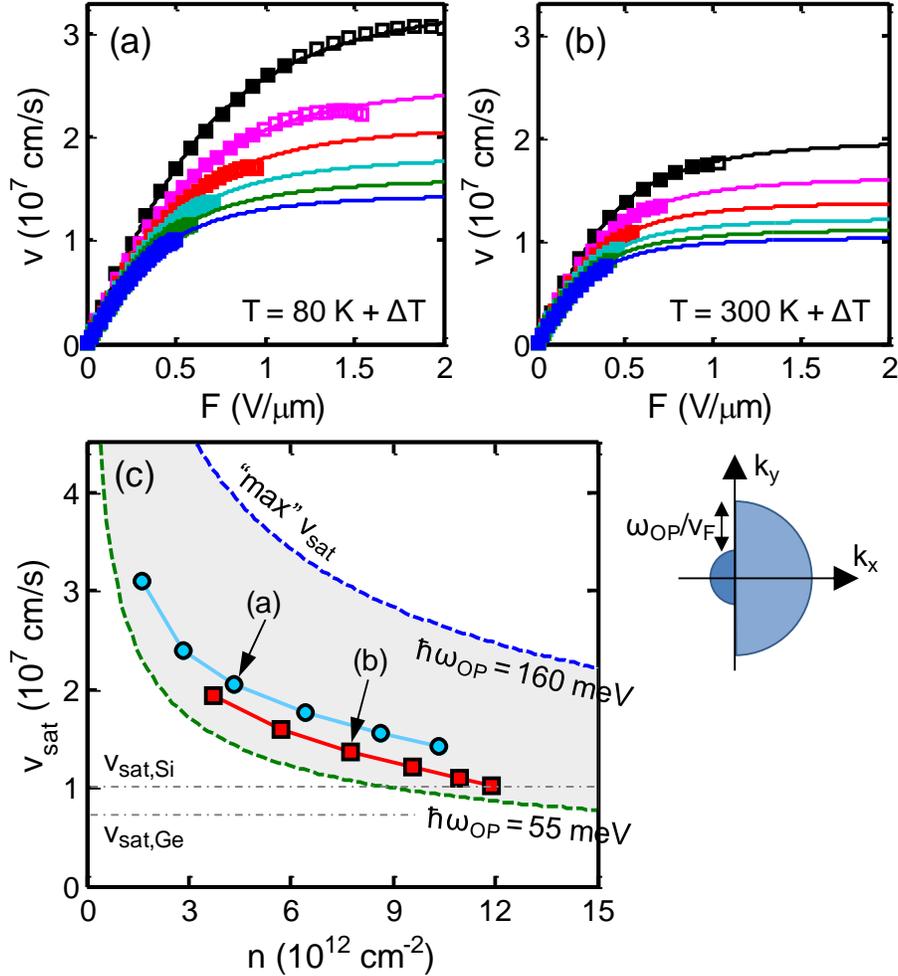

**FIG. 3** Electron saturation velocity. **(a)** Background temperature $T_0 = 80$ K with $V_{G0} = 10.5$–60.5 V, and **(b)** $T_0 = 300$ K with $V_{G0} = 23.5$–73.5 V (in 10 V steps from top to bottom). Squares represent data, lines are empirical fits with Eq. (6); open squares have $\Delta T > 200$ K from Joule heating and were not used for fit. Changing fitting criteria results in ±8% uncertainty. **(c)** Saturation velocity vs. electron density at $F = 2$ V/µm. Side panel shows carrier distribution assumed for analytic model. Dashed lines show Eq. (7) with $\hbar\omega_{OP} = 55$ meV (SiO$_2$) and 160 meV (graphene), suggesting the maximum $v_{sat}$ that could be achieved in graphene. Theoretical studies[23] predict comparable $v_{sat} \approx 2$-5×10$^7$ cm/s in carbon nanotubes. Electron $v_{sat}$ for Si and Ge are appreciably lower, but largely independent of carrier density.[21]

**Supporting Online Materials** for "Mobility and Saturation Velocity in Graphene on SiO₂" by V. E. Dorgan, M.-H. Bae, and E. Pop, University of Illinois, Urbana-Champaign, U.S.A. (2010)

**1. Graphene Device Fabrication:** We deposit graphene from mechanical exfoliation of natural graphite onto SiO₂ (dry thermally grown, thickness $t_{ox} = 300$ nm) with an n+ doped ($2.5 \times 10^{19}$ cm⁻³) Si substrate, which also serves as the back-gate (Fig. 1a). Electron beam lithography is used to define the four Cr/Pd (0.5/40 nm) electrodes, with inner voltage probes much narrower (~300 nm) than the graphene channel dimensions ($W = 7$ μm, $L = 4$ μm), to provide minimally invasive contacts. An additional lithography step followed by an oxygen plasma etch (75 W, 0.1 Torr for 15 s) creates the graphene channel.[1]

**2. Raman Spectroscopy:** After device fabrication, Raman spectroscopy is used to confirm that the graphene flake is indeed monolayer graphene. The Raman 2D peak of monolayer graphene exhibits a single Lorentzian line shape.[2] In this study, Raman spectra were obtained using a Renishaw Raman spectrometer with a 633 nm laser excitation (power at the object: 10 mW) and a 50× in-air objective. Figure S1a shows the Raman spectrum obtained from the fabricated graphene device shown in Fig. 1a. The single Lorentzian fit to the 2D peak in Fig. S1 confirms that this sample is monolayer graphene, as does the approximate ratio (1:2) of the G to 2D peaks.

**3. Extracting Impurity Density:** The charged impurity density at the SiO₂ surface is determined based on the approach discussed in Ref. [3] and given by $n_{imp} = BC_{ox}|d\sigma/dV_{G0}|^{-1}$ where $B = 5 \times 10^{15}$ cm⁻² is a constant determined by the screened Coulomb potential in the random phase approximation (RPA)[4] and $d\sigma/dV_{G0}$ is the slope of the low-field conductivity $\sigma = (L/W)(I_{14}/V_{23})$.

The slope is determined by a linear fit to $\sigma$ over a $\Delta V_{G0} = 2$ V interval around the maximum of $|d\sigma/dV_{G0}|$ as shown in Fig. S1b. The value of $n_{imp}$ used here is based on a conductivity measurement at 80 K, where mobility is limited by Coulomb scattering.[5] From Eq. (10) of Ref. 6, we determine $n^* \approx 0.297 n_{imp} \approx 2.63 \times 10^{11}$ cm⁻² in our sample (averaged over the electron- and hole-doped regimes), where $n^*$ is the residual carrier puddle density representing the width of the minimum conductivity plateau. From $n^*$, we obtain $\Delta \approx \hbar v_F (\pi n^*)^{1/2} \approx 59$ meV and $\Delta V_0 \approx q n^*/C_{ox} \approx 3.66$ V in order to model the spatially inhomogeneous electrostatic potential (main text).

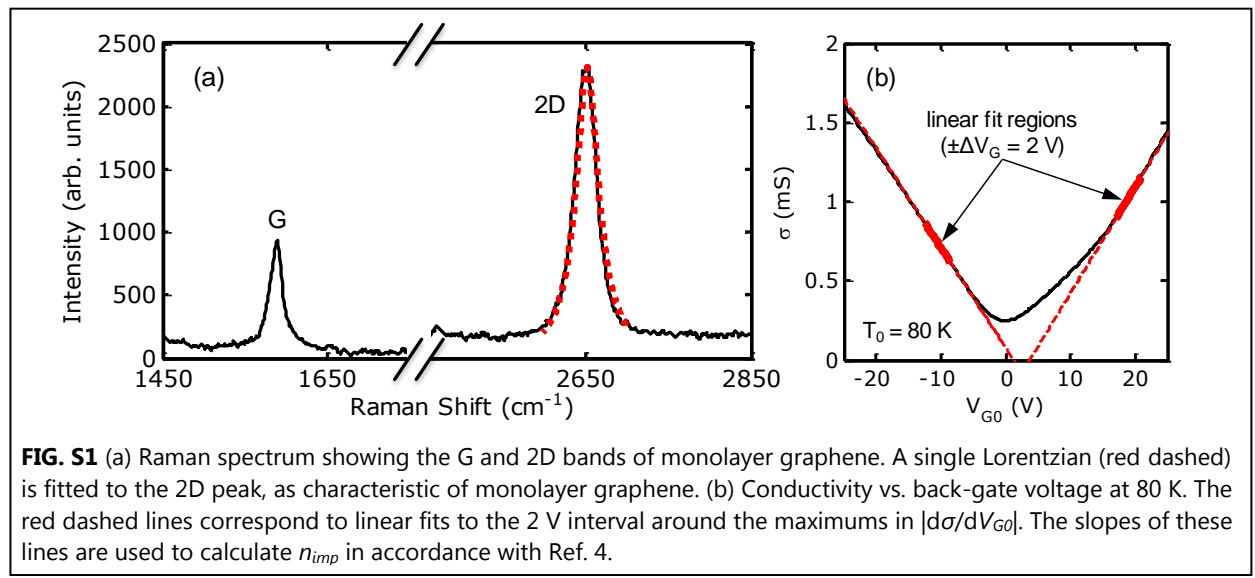

**FIG. S1** (a) Raman spectrum showing the G and 2D bands of monolayer graphene. A single Lorentzian (red dashed) is fitted to the 2D peak, as characteristic of monolayer graphene. (b) Conductivity vs. back-gate voltage at 80 K. The red dashed lines correspond to linear fits to the 2 V interval around the maximums in $|d\sigma/dV_{G0}|$. The slopes of these lines are used to calculate $n_{imp}$ in accordance with Ref. 4.

**4. Hole Mobility and Uncertainty:** As in the main text, the mobility for the hole-doped regime ($V_{G0} < 0$) is obtained and shown in Fig. S2a. Given the symmetric energy bands in graphene, the hole mobility here is similar to the electron mobility from Fig. 2a. Only one discrepancy exists for the 300 K data set, which does not show the typical mobility peak vs. carrier density as in all other measurements. This is most likely due to the calculated hole density underestimating the actual value in the device, probably due to a spatial inhomogeneity of the "puddle" regime that is not accurately predicted by the simple $\pm\Delta$ potential model. We note the mobility shape recovers at higher temperatures (350-500 K), as thermal smearing washes out such inhomogeneities.

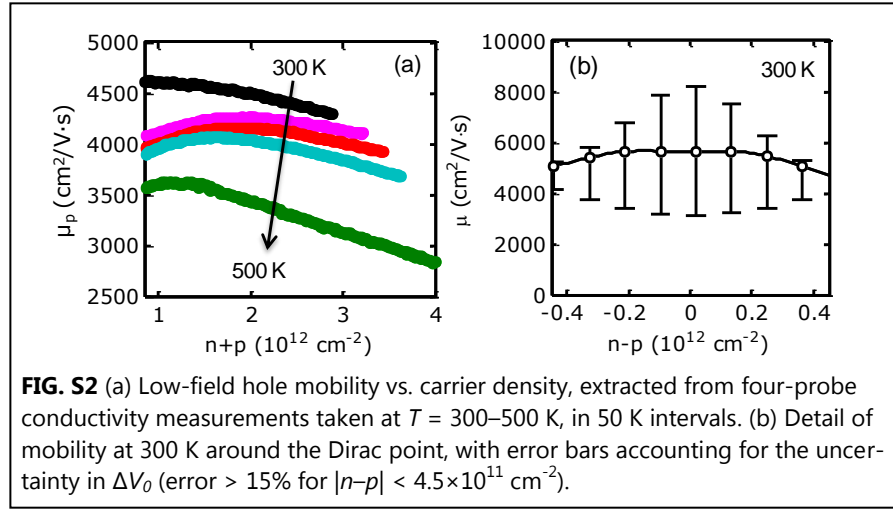

**FIG. S2** (a) Low-field hole mobility vs. carrier density, extracted from four-probe conductivity measurements taken at $T = 300$–$500$ K, in 50 K intervals. (b) Detail of mobility at 300 K around the Dirac point, with error bars accounting for the uncertainty in $\Delta V_0$ (error > 15% for $|n-p| < 4.5\times10^{11}$ cm$^{-2}$).

To understand additional uncertainty associated with this method, Ref. 6 noted that for the extracted $n_{imp}$ the predicted plateau width is approximate within a factor of two. This leads to uncertainty in determining $\Delta V_0$, and thus uncertainty in the charge density and extracted mobility values. However, this uncertainty is only notable around the Dirac point, where the potential ripple contributes to the total carrier density. This limits the "confidence region" in Figs. 2a and S2a, with charge density shown only $>0.85\times10^{12}$ cm$^{-2}$. In Fig. S2b we estimate the mobility uncertainty at lower charge densities around the Dirac point, such that the upper and lower bounds result from a potential ripple of $\Delta V_0/2$ and $2\Delta V_0$ respectively.

**5. Additional Sources of Uncertainty:** Our devices use inner voltage probes that span the width of the graphene sheet. The advantage of such probes is that they sample the potential uniformly across the entire graphene width, unlike edge-probes which may lead to potential non-uniformity particularly at high fields. However, full-width probes themselves introduce a few uncertainties in our measurements, which are minimized through careful design as described here.

One challenge may be that, even under four-probe measurements, the current flowing in the graphene sheet could enter the edge of a voltage probe and partially travel in the metal. To minimize this effect, we used very narrow inner probes (300 nm), narrower than the typical charge transfer length between graphene and metal contacts (~1 μm).[7] In addition, we employed very "long" devices, with $L = 4$ μm or longer between the inner probes. Thus, the resistance of the graphene between terminals 2-3 is much greater than both the resistance of the graphene under the metal contact and that across the narrow metal contact itself.

A second challenge is that of work function mismatch between the Cr/Pd electrodes and that of the graphene nearby. This leads to a potential and charge non-uniformity in the graphene near the contacts. Recent theoretical studies[8] have calculated a potential decay length of ~20 nm induced by Pd contacts on graphene. Experimental photocurrent studies have estimated that doping from Ti/Pd/Au contacts can extend up to 0.2-0.3 μm into the graphene channel,[9] although the study had a spatial resolution of 0.15 μm. In either case, the potential and charge disturbance is much shorter than the total channel length of our device (4 μm). We estimate at most a ~10% contribution to the resistance from charge transfer at our metal contacts, an error comparable to the ±8% from fitting the high-field velocity data in Fig. 3. The error is likely to become smaller at higher change densities (>2×10$^{12}$ cm$^{-2}$) where the graphene charge more strongly screens the contact potential, and at high fields where the graphene channel becomes more resistive itself.

A third challenge is that of temperature non-uniformity around the inner voltage probes, which may act as local heat sinks. The thermal resistance "looking into" the metal voltage probes can be estimated as $\mathcal{R}_c = L_T/\kappa_m A$, where the thermal healing length $L_T = (t_m t_{ox} \kappa_m/\kappa_{ox})^{1/2} = 685$ nm, $t_m = 40$ nm is the metal thickness, $t_{ox} = 300$ nm is the SiO$_2$ thickness, $\kappa_m \approx 50$ Wm$^{-1}$K$^{-1}$ is the Pd metal thermal conductivity, $\kappa_{ox} \approx 1.3$ Wm$^{-1}$K$^{-1}$ is the oxide thermal conductivity (at 300 K), and $A=t_m W_c$ is the cross-sectional area of the contact with $W_c = 300$ nm. We obtain $\mathcal{R}_c$ ~10$^6$ K/W based on the device geometry here (primarily due to the narrow inner contacts being only ~300 nm wide) which is about two orders of magnitude greater than the thermal resistance for heat sinking from the large graphene sheet through the oxide (Fig. 1a and Eq. (5)), $\mathcal{R}_{th} \approx 10^4$ K/W at 300 K. Thus, heat flow from the inner metal contacts is negligible.

**6. Temperature Dependence of Thermal Resistance:** Self-heating during high-field measurements is determined by Eq. (4), which raises the temperature of the graphene device and of the underlying SiO$_2$ as a function of input power during the measurement. The thermal resistance $\mathcal{R}_{th}$ depends on temperature through $\kappa_{ox}$ (thermal conductivity of SiO$_2$) and $\kappa_{Si}$ (thermal conductivity of the doped Si wafer). These can be written as $\kappa_{ox} = \ln(T_{ox}^{0.52}) - 1.687$ and $\kappa_{Si} = 2.4\times10^4/T_0$ by simple fitting to the experimental data in Refs. 10 and 11 respectively. Based on the thermal resistance model, we estimate the average oxide temperature as $T_{ox} = (T_0+T)/2$ and the temperature of the silicon substrate as the background temperature $T_0$. This allows a simple iterative method for calculating the graphene temperature rise ($\Delta T$) during measurements.

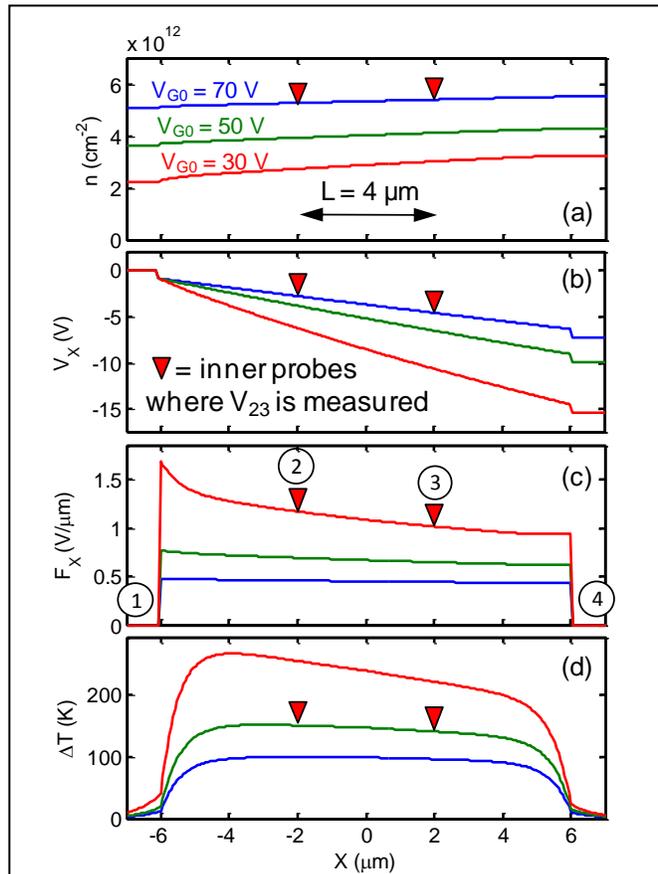

**FIG. S3** Using the finite-element simulations discussed in Ref. 12, we plot (a) electron density $n$, (b) channel potential $V_x$, (c) electric field $F$, and (d) temperature increase $\Delta T$ across the length of the device for back-gate voltages of $V_{G0} = 30$ V (red), 50 V (green), and 70 V (blue). The simple assumptions of relatively uniform charge density and constant electric field (main text) are acceptable if ambipolar transport is avoided, and high-field measurements are done at average charge densities >2×10$^{12}$ cm$^{-2}$. In addition, we note that high-field non-uniformities may still occur at the outer electrodes (1 and 4) but not in the relevant channel portion between the inner electrodes (2 and 3).

**7. Electrostatics in the High-Field Regime:** During high-field measurements, the carrier density and temperature can vary across the device.[12] As mentioned in the main text, we minimize this effect by carefully avoiding ambipolar transport and generally restricting our samples to average carrier density >$2\times10^{12}$ cm$^{-2}$ at high fields. In addition, here we follow Ref. 12 to fully model this regime as shown in Fig. S3. Based on these simulation results, we confirm that it is valid to assume a constant (average) electric field and relatively uniform charge density across the active region of the device between the inner electrodes, if the biasing scheme mentioned above is followed. The average carrier density in the channel is simply given by the average of the densities at electrodes 2 and 3 (as labeled in Fig. 1a), while the average field is $F = (V_2-V_3)/L$.

**8. Original I-V Data:** For completeness, we include here in Fig. S4 the original I-V data taken at high field to extract the drift velocity in Fig. 3 of the main text.

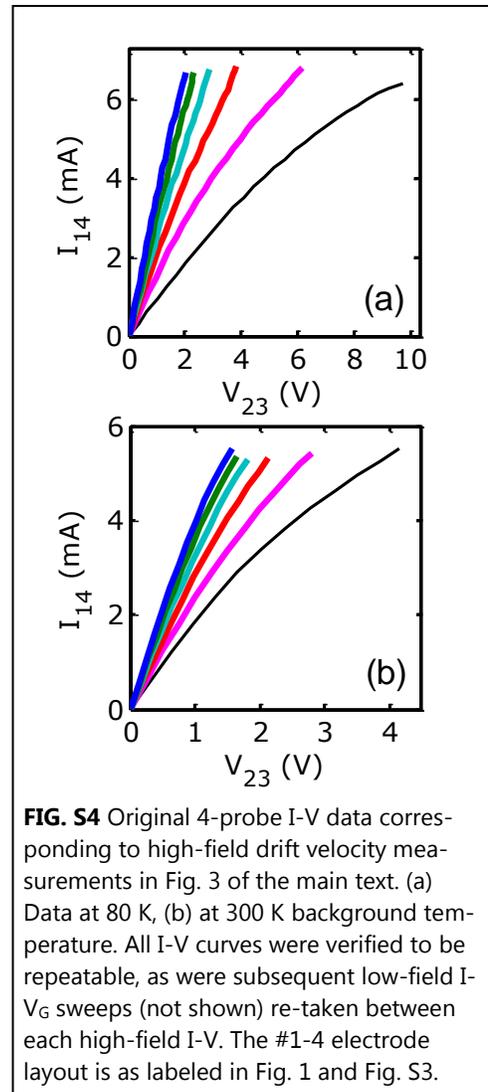

**FIG. S4** Original 4-probe I-V data corresponding to high-field drift velocity measurements in Fig. 3 of the main text. (a) Data at 80 K, (b) at 300 K background temperature. All I-V curves were verified to be repeatable, as were subsequent low-field I-$V_G$ sweeps (not shown) re-taken between each high-field I-V. The #1-4 electrode layout is as labeled in Fig. 1 and Fig. S3.